# Signatures of Steady Heating in Time Lag Analysis of Coronal Emission

Nicholeen M. Viall and James A. Klimchuk

NASA Goddard Space Flight Center, Solar Physics Laboratory, Greenbelt, MD 20706, USA

**Abstract**

Among the many ways of investigating coronal heating, the time lag method of Viall & Klimchuk (2012) is becoming increasingly prevalent as an analysis technique complementary to those traditionally used. The time lag method cross correlates light curves at a given spatial location obtained in spectral bands that sample different temperature plasmas. It has been used most extensively with data from the Atmospheric Imaging Assembly on the Solar Dynamics Observatory. We have previously applied the time lag method to entire active regions and surrounding quiet Sun and create maps of the results (Viall & Klimchuk 2012; Viall & Klimchuk 2015). We find that the majority of time lags are consistent with the cooling of coronal plasma that has been impulsively heated. Additionally, a significant fraction of the map area has a time lag of zero. This does not indicate a lack of variability. Rather, strong variability must be present, and it must occur in phase in the different channels. We have shown previously that these zero time lags are consistent with the transition region response to coronal nanoflares (Viall & Klimchuk 2015; Bradshaw & Viall 2016), but other explanations are possible. A common misconception is that the zero time lag indicates steady emission resulting from steady heating. Using simulated and observed light curves, we demonstrate here that highly correlated light curves at zero time lag are *not* compatible with equilibrium solutions. Such light curves can only be created by evolution.

## I. Introduction

Understanding how the solar corona is heated to temperatures of 1 MK and greater is one of the outstanding problems of space science (e.g., Klimchuk 2015). There are many observational approaches to solving this so-called 'solar coronal heating problem', including the measurement of differential emission measure (DEM) at a given location and time, and analysis of coronal loop light curves in active regions (see review by Reale 2014). A complementary method for addressing the coronal heating problem is the time lag method of Viall & Klimchuk (2012). In this method, light curves of different observing channels with unique temperature sensitivity are computed at individual pixels and cross correlated with one another. The cross correlation is repeated at different temporal offsets, typically ranging from -2 hours to + 2 hours. The time lag between two light curves is defined as the time offset at which the highest cross correlation value is reached.

The time lag method combines several advantages of other methods. Like DEM analysis, usually performed with spectrometer data, it can address all of the emission along the line of sight and does not require the subtraction of a background emission that loop studies require. Like loop studies, the time lag method takes advantage of the temporal coverage and large field of view of EUV imagers. Additionally, since every pixel is treated individually, large statistics can be compiled, and the variation in properties within and between different features can be addressed. The time lag method is especially well suited to the Atmospheric Imaging Assembly (AIA) on board the Solar Dynamics Observatory (SDO) (Lemen et al. 2011; Boerner et al. 2011), with its continuous observation of the full Sun at high cadence in 6 EUV channels. By comparing different pairs of the channels, 15 unique time lag maps can be produced. Together, they provide a very stringent test for any coronal heating model. This is a very powerful diagnostic tool, especially when combined with complementary information from other techniques such as the measurement of DEM slopes (e.g., Bradshaw et al. 2012; Warren et al. 2012).

The time lag method is becoming increasingly used by the solar community (e.g. Lionello et al. 2016; Tajfirouze et al. 2016; Bradshaw & Viall 2016; Froment et al. 2015; Viall & Klimchuk 2012, 2013, 2015), and therefore a full explanation of the nuances of the measurements is necessary. The key to the time lag technique is that it identifies variability in light curves. Such variability can have many sources, but plasma evolution associated with sudden bursts of heating seems to be especially important. When a nanoflare (defined as an impulsive burst of energy release) occurs, the heating phase contributes very little to the total emission, and it is the subsequent cooling that dominates the light curves (Bradshaw & Klimchuk 2011). The signature in an instrument such as SDO/AIA is that light curves in the hottest channel will rise and fall first, followed later in time by a rise and fall in the next coolest channel and so on (Viall & Klimchuk 2011, 2012). The time lag technique will identify this cooling when it is present. Viall & Klimchuk (2013) demonstrated that the time lag is easily recovered even when thousands of out-of-phase events contribute to the integrated emission along the line of sight and the resulting composite light curves have only small fluctuations. Indeed, Viall & Klimchuk (2012, 2015) found that the majority of time lags are consistent with cooling following an impulsive heating event.

In addition to the time lags that are consistent with cooling coronal plasma, we found a significant number of time lags of zero, or light curves with variance that are highly correlated at no temporal offset. We have shown that these zero time lags are consistent with the transition region response to coronal nanoflares (Viall & Klimchuk 2015; Bradshaw & Viall 2016). We define the top of the transition region as the physical location where thermal conduction changes from an energy sink (above) to an energy source (below) (Vesecky et al. 1979). When defined this way, the shape of DEM(T) in the transition region changes only minimally, as described in Viall & Klimchuk (2015), and shown with an analytic derivation of DEM(T) in the appendix of Klimchuk et al. (2008). Though the amplitude of DEM(T) will move up and down as a flux tube evolves, the temperature dependence is largely unaffected. All layers (temperatures) of the transition region respond in unison, so the resulting emission in different channels brightens and fades together and there is near zero temporal offset between light curves. Additionally, Viall & Klimchuk (2015) showed that the transition region will have significant emission in all of the SDO/AIA channels when there is an impulsive coronal heating event. These conclusions were confirmed by the result that far fewer zero time lags are present in limb observations, at altitudes above where the transition region exists.

While impulsive heating is a viable explanation for zero time lags for lines of sight intersecting the transition region, other possibilities also exist, which we discuss in section III. One explanation that can be easily ruled out is static equilibrium resulting from steady heating. It is a misconception that steady emission produces zero time lag (e.g., Lionello et al. 2016). This may seem counterintuitive, since there is of course no time scale associated with steady emission. A crucial point is that the absence of variability is very different from variability that is strongly correlated at zero time lag.

No emission is truly steady, however. Even if the emitting plasma is not evolving, there are random intensity fluctuations associated with statistical counting noise. Because the fluctuations have no physical correlation between channels, there is no preferred time lag. All time lags occur with equal likelihood. Furthermore, the strength of the correlation, as measured by the correlation coefficient, is very weak. As we show below, steady emission from steady heating then produces maps of random time lags with low cross correlations. On the other hand, if a multi-pixel region of near zero time lag is present in a time lag map, it is a clear indication of a dynamic situation. The emission must have substantial variability in each pixel, and it must be evolving in unison at the different observed temperatures. We now demonstrate these points with simulated observations based on simple models.

## II. Results

In this paper we model 12 hours of steady emission in the 211 and 171 Å channels of SDO/AIA and apply the time lag test to the resulting light curves. We create $1.6 \times 10^5$ instances (pixels) of emission in 211 and 171 Å with Poisson noise fluctuations about a constant value. For 211 Å, we used count rates of 1300 cts s$^{-1}$ and for 171 Å we used 1000 cts s$^{-1}$, chosen based on the mean intensities in the core of active region 11082 observed on 2010 June 19 (from Viall & Klimchuk 2012), and well within the typical observed values in this active region. We will show below that the results are independent of count rate. The light curves have a 12 second cadence, to match the AIA observations. At each pixel we cross correlate the 211 Å light curve with the 171Å light curve as a function of temporal offset, similar to the single instance shown in Viall & Klimchuk (2013). We test for temporal offsets of up to +/- 2 hours. For each pixel, we record the time lag and the cross correlation value of that time lag.

In Figure 1, we show a histogram of the time lags computed between the channels. On the upper left, we show those computed for the noise model. On the upper right, we show the results calculated in Viall & Klimchuk (2012) for active region 11082 using observations from 0-12 UT 2010 June 19. The y-axis is the same for both plots and is the logarithm of the number of pixels in that histogram bin. The x-axis is the time lag value, from the possible +/- 7200 seconds tested, and each histogram bin is 24 seconds wide. We expect the spectrum of fluctuations to be independent of the noise level, so the result should not depend on what we choose for the average count rate. Nonetheless, we have performed additional simulations with ten times larger (lower left panel) and 10 times smaller (lower right panel) average count rates and verified that the result is the same.

In the case of noise, the time lags are nearly evenly distributed across all of the values tested. Negative time lags were found in 50.0%, 50.0%, and 49.9% of the pixels for the base model, the model with ten times the count rates, and the model with 1/10 of the count rates, respectively. Zero time lag (between ±12 s) was found in 0.2%, 0.1%, and 0.1 % of the pixels in the noise models, and a positive time lag was found in 49.8%, 49.9%, and 50.0% of the pixels in the noise models, all indicated in Figure 1. Though the distribution is symmetric about zero, it is not perfectly flat: there is a slight preference towards shorter time scales. This is a small effect and is due to how the method treats finite time series. When cross correlating the 12 hour light curves at zero temporal offset, the entire 12 hours of both time series is used. When cross correlating the 12 hour light curves at a 2 hour temporal offset, the first two hours of the first time series and the last two hours of the second time series are omitted, and the cross correlation is performed on only 10 hours of data.

For the observed active region, for this pair of channels, positive time lags are consistent with cooling from ~ 2 MK to less than 1 MK, and are the expected signature from cooling of impulsively heated plasma. These dominate the field of view (Fig. 5 of Viall & Klimchuk 2012), in the main body of the active region, and comprise 56% of all of the pixels. This is especially significant considering that the field of view included the transition region/moss of the active region, fan loops, and surrounding quiet Sun. The number of zero time lags were also large, comprising 23% of the pixels, with negative time lags comprising 21% of the field of view. Thus, at least 79% of the imaged plasma is behaving in a way consistent with impulsive heating in the corona.

In Figure 2, we show a histogram of the cross correlation values of the time lags shown in Figure 1, with the case of noise plus steady emission at three different count rates on the upper left, lower left, and lower right, and the case of the observations on the upper right. The y-axis is the number of pixels, and the x-axis is the cross correlation value, in 0.01 sized bins. When noise is the source of variability, the cross correlation value of the time lags is very low: between 0.0 and 0.01 for 100% of the time lags computed in all of the noise models. In contrast, in the observed active region, 89% of pixels exhibited cross correlations greater than 0.1. The remaining 11% exhibited negative cross correlations (i.e. they were anti-correlated) or very low correlations consistent with noise.

## III. Discussion

Two things are clear from Figures 1 and 2. If the emission in a multi-pixel region of the Sun were steady or otherwise dominated by noise, then the measured time lags would have a random scatter over the full range of values allowed by the search window (±2 hours in our case). The region would *not* be dominated by time lags of zero, and it would not appear uniform in a time lag map. Second, the cross correlation values associated with the time lags would be very small (< 0.1 magnitude). Regions with these properties are uncommon in actual solar data, and typically occur in pixels with extremely low count rates (see e.g. appendix of Viall & Klimchuk 2012) where noise fluctuations dominate any physical variations. Note that since the result is independent of count rate and is a fundamental property of noise fluctuations, it applies equally to all locations within the active region, as well as other locations on the Sun, where the temperature and ratio of the count rates are different. Our results also apply to other channels and to other EUV imaging instruments.

We conclude that dynamic plasma is present along the overwhelming majority of lines of sight on the Sun. We do not, however, rule out the possible coexistence of static plasma along these same lines of sight. The important question of the relative proportions of dynamic and static plasma has yet to be answered. We have devised a method for future studies for finding the answer that involves a combination of time lags, cross correlation coefficients, mean intensities, and intensity fluctuation envelopes. For now, we can say that the fraction of dynamic plasma along most lines of sight is not small.

We have already emphasized that the transition region response to a coronal nanoflare will produce a time lag of zero. This explains why many regions of zero time lag are well correlated with moss (Viall & Klimchuk 2015). Viall & Klimchuk (2015) showed light curves computed from a model transition region connected to an impulsively heated corona. The 211 and 171 Å light curves and time lag from Figure 5 of Viall & Klimchuk (2015) are reproduced in the bottom two left panels here for easy comparison. The light curves are a model of a single AIA pixel imaging the transition region of many thousands of unresolved flux tubes undergoing heating from random nanoflares. The modeled light curves are variable and highly correlated (cross correlation value of 0.95) with a time lag of zero. In the top left and middle panels of Figure 3, we show light curves in 211 and 171 Å for two examples of pixels in the moss of NOAA active region 11082 on 2010 June 19. Each example was chosen from regions identified as moss (using magnetic field strength contours of ±100 G to indicate flux tube foot points) in Viall & Klimchuk (2015), and were found to have a zero time lag in Viall & Klimchuk (2012). The left panel shows an example taken from the footpoints in the upper left portion of the active region, while the middle panel shows one taken from the footpoints in the lower right portion. Note that in the middle panel the count rates are very different between 211 and 171 Å and so they have different y axes. The bottom left and middle panels show the corresponding time lags computed for those pairs of light curves. In both cases, the 211 and 171Å light curves are highly variable, and their variations generally occur at the same time - i.e. with zero lag. The time lag is the time offset at which the cross correlation value peaks, indicated with a black dot. The light curves are highly correlated with each other, with both examples producing time lags of zero, at well over 0.8 cross correlation value.

In the right panels, we show light curves and the corresponding time lag calculation for one of the rare examples where the 211 and 171 Å light curves are uncorrelated. This example is from the lower right side of the same active region, near the footpoints, but not located in the moss. In the lower right two panels we show the 211 Å and 171 Å light curves and associated time lag from a model of steady emission and noise (following Figures 3 and 4 of Viall & Klimchuk, 2013). In the case of the model, a random time lag of -3420 seconds is computed with an extremely low, uncorrelated cross correlation value of 0.07. In the observed case, there are substantial count rates, and both light curves have substantial variability well above that expected from counting statistics, so the observed uncorrelated case is also inconsistent with steady emission from steady heating. In this example, the variations in 211 and 171 Å are largely uncorrelated, producing a cross correlation value of only 0.02. In this case, we conclude that there are two independent populations of plasma, one around 0.8 MK (the peak sensitivity of the 171 Å channel) and one around 2 MK (the peak sensitivity of the 211 Å), with neither the 0.8 MK plasma heating into the 211 Å sensitivity, nor the 2 MK plasma cooling into the 171 Å sensitivity, for this period of time.

The cooling coronal plasma associated with nanoflares can also produce a zero time lag if the nanoflares repeat with sufficient frequency that the plasma cools only partially before being reheated (Bradshaw & Viall 2016). For this to be the case, the plasma must cool into, but not cool entirely through the range of peak sensitivity of the channel pair. Lastly, zero time lags can also result from motion within the image plane, either due to large amplitude waves or to evolution of the magnetic field. Consistent with this picture, Uritsky et al. (2013) found plasma density motion associated with slow mode waves in the regions of the fan loops where Viall & Klimchuk (2012) measured zero time lag pixels. When plasma is carried into or out of a pixel, the emission seen in that pixel will be variable. The plasma need not be heating or cooling. If the plasma is multi-thermal, or if it is isothermal but in the range of significant sensitivity for both observing channels, then the light curves will evolve in unison. They will be strongly correlated at zero time lag.

## IV. Conclusion

In this paper, we have shown that the vast majority of the time lags - including the zero time lags - found in time lag maps presented in Viall & Klimchuk (2012, 2015) cannot be explained by steady emission. We modeled steady emission produced by steady heating with the addition of Poisson noise and computed time lags on $1.6 \times 10^5$ simulated pixels. We found that steady emission will result in a random distribution of time lags. All of the tested time lags will be randomly sampled, there will be no coherence of time lag along physical structures, and there will be no coherence of time lags between different sets of time lag maps produced between different channel pairs. Steady emission produces a zero time lag in approximately 0.2% of pixels. In all cases of steady emission, the cross correlation value was less than 0.1. Importantly, any time lags computed with cross correlation values greater than ~ 0.1 indicate highly dynamic plasma, and the zero time lags are no exception. We demonstrated this using count rates and properties of the 211 Å and 171 Å SDO/AIA filters, but the result is independent of count rate, and these conclusion hold true for the time lag method applied to the combinations with the other 4 channels in SDO/AIA, and more broadly, to any measurement of emission in a narrow temperature range.

Acknowledgements

This research was supported by a NASA GI grant.

Figures

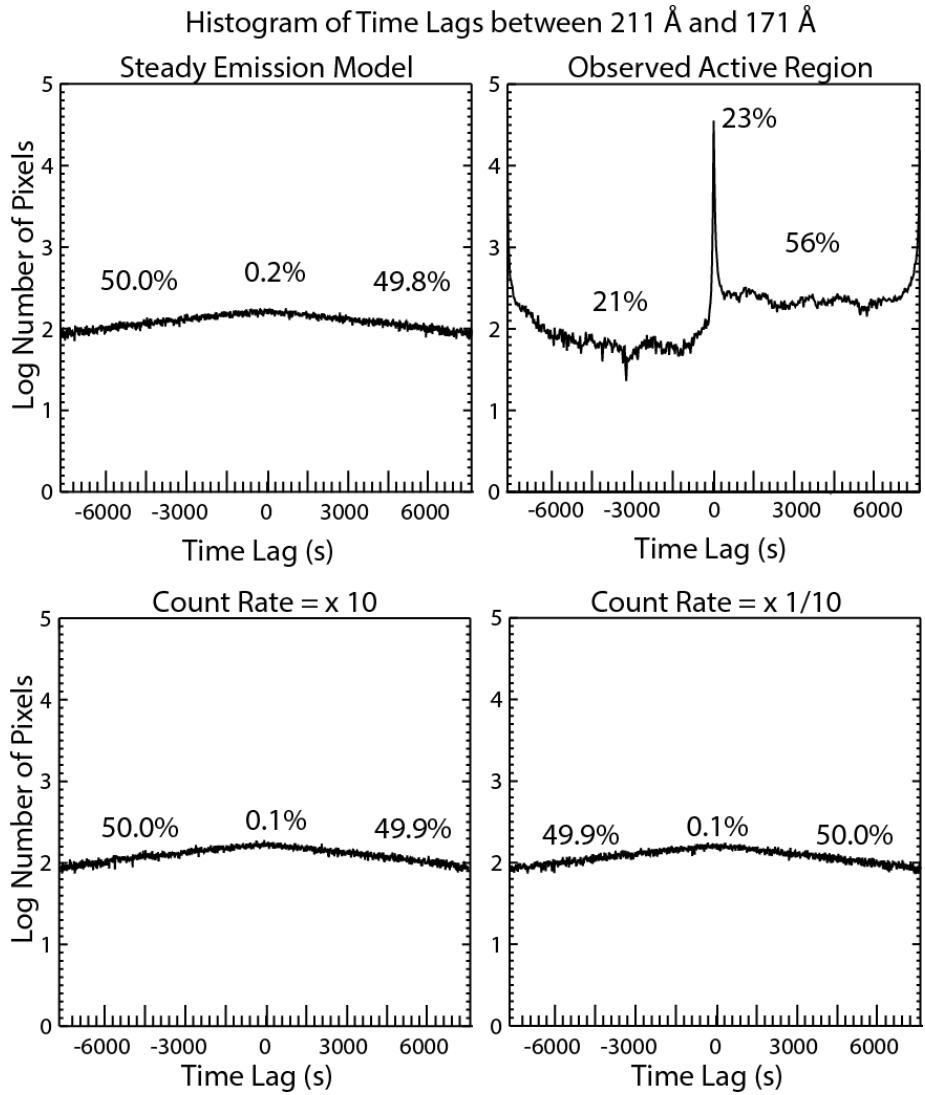

Figure 1. Histograms of time lags calculated for the steady emission model (upper left) and for the observed active region (upper right). The y-axis is the logarithm of the number of pixels, and the x-axis are the time lags tested in 24-second wide bins. Percent of negative, zero (between ±12 s), and positive time lag pixels are listed in each plot. Lower panels show model results with ten times the count rate (left) and 1/10 the count rate (right) of the upper left panel.

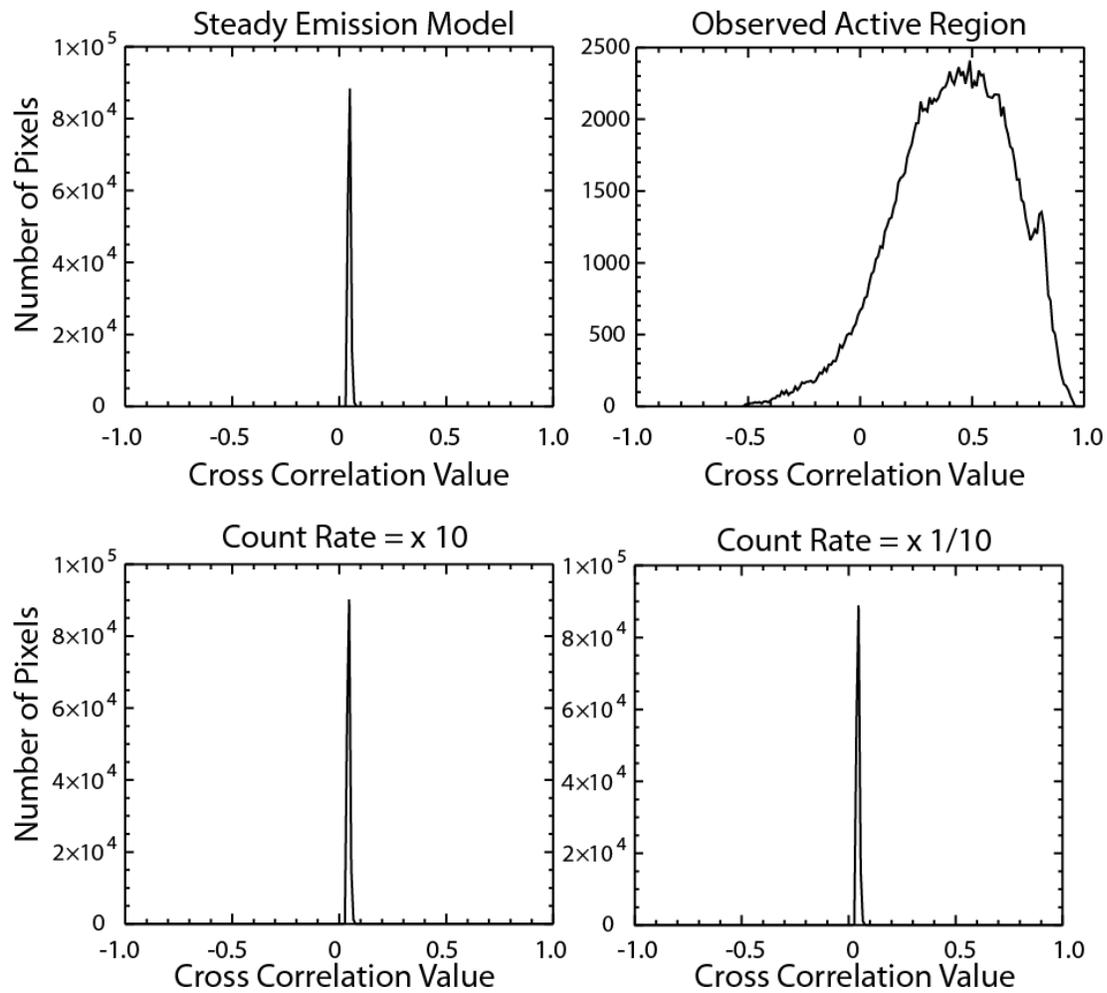

Figure 2. Histograms of the cross correlation values of the time lags shown in Figure 1. Upper left is the steady emission model and upper right is the observed active region. The y-axis is the number of pixels, and the x-axis is the cross correlation value, in 0.01 wide bins. Lower panels show model results with ten times the count rate (left) and 1/10 the count rate (right) of the upper left panel.

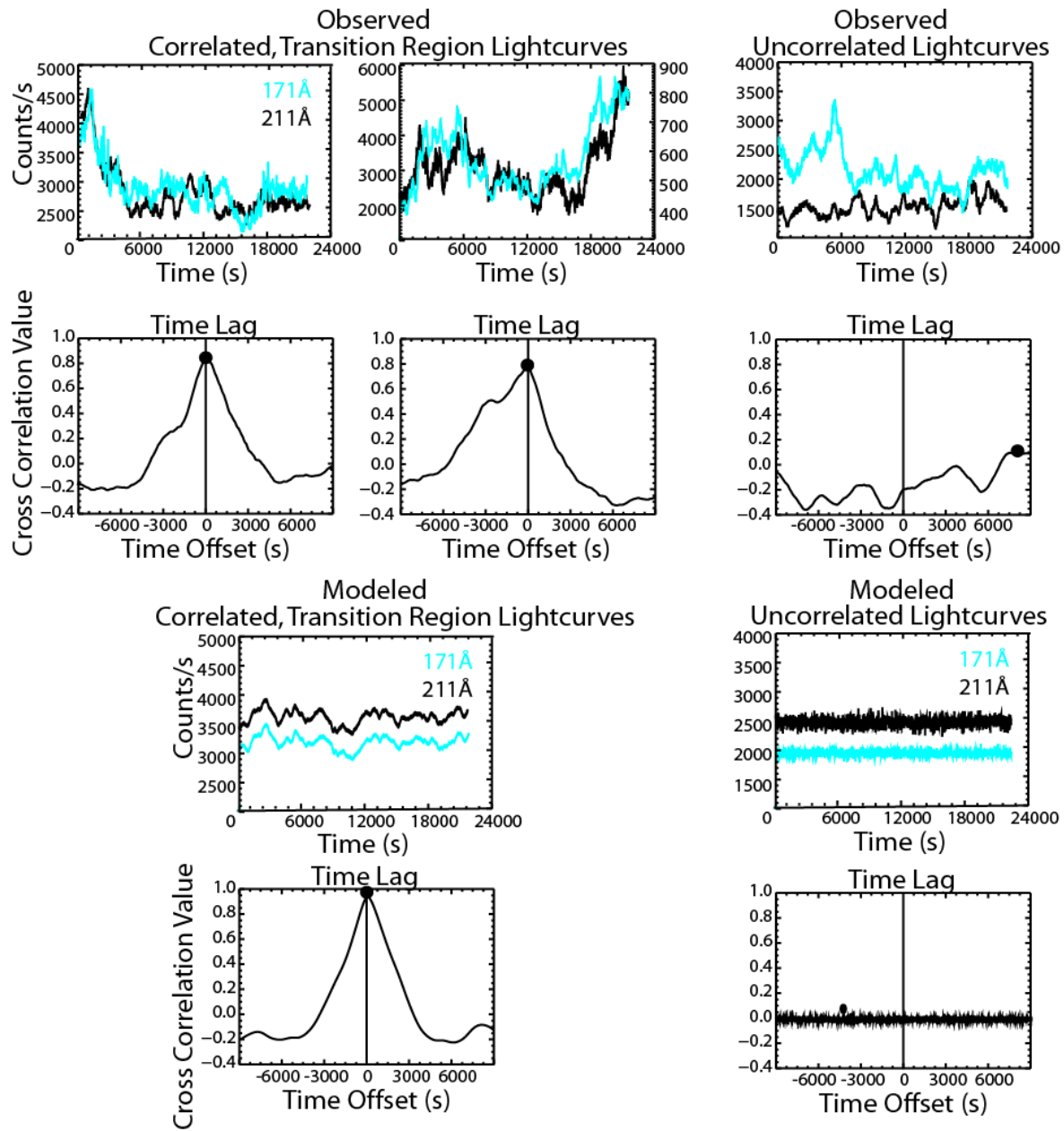

Figure 3. Top panels show observed light curves from three locations in NOAA AR 11082 in 211 (black) and 171 (cyan) Å. The second row of panels show corresponding cross correlation values as a function of time offset. The time lag is indicated with a black dot. The left two are locations in the moss which are highly correlated at a time lag of zero. The right location shows an instance where the light curves are largely uncorrelated. The bottom panels are from a model of transition region light curves (left), and steady emission with noise (right) and their time lags underneath, adapted from Viall & Klimchuk (2015) and Viall & Klimchuk (2013).